\documentclass[conference]{IEEEtran}

\usepackage{graphicx}
\usepackage{subcaption}  
\usepackage{cite}
\usepackage{amsmath}
\usepackage{amssymb}

\usepackage{makecell}
\usepackage{tikz}
\usetikzlibrary{positioning}
\usepackage{xspace}

\usepackage[hidelinks]{hyperref}
\usepackage[capitalize,noabbrev]{cleveref}

\def \alg{\texttt{DEFINED}\xspace}

\ifodd 1
\newcommand{\congr}[1]{{\color{blue}#1}}
\else
\newcommand{\congr}[1]{#}
\fi

\ifodd 1
\newcommand{\congc}[1]{{\color{red}(Cong: #1)}}
\else
\newcommand{\congc}[1]{}
\fi

\title{Decision Feedback In-Context Symbol Detection over Block-Fading Channels}
\author{
    Li Fan\textsuperscript{1}, Jing Yang\textsuperscript{2}, and Cong Shen\textsuperscript{1} \\
    \textsuperscript{1}Charles L. Brown Department of Electrical and Computer Engineering, University of Virginia, USA \\
    \textsuperscript{2}Department of Electrical Engineering, The Pennsylvania State University, USA \\
    E-mails: {\{lf2by,cong\}@virginia.edu}, {yangjing@psu.edu}
}

\begin{document}

\maketitle

\begin{abstract}
Pre-trained Transformers, through in-context learning (ICL), have demonstrated exceptional capabilities to adapt to new tasks using example prompts \textit{without model update}. Transformer-based wireless receivers, where prompts consist of the pilot data in the form of transmitted and received signal pairs, have shown high estimation accuracy when pilot data are abundant. However, pilot information is often costly and limited in practice. In this work, we propose the \underline{DE}cision \underline{F}eedback \underline{IN}-Cont\underline{E}xt \underline{D}etection (\alg) solution as a new wireless receiver design, which bypasses channel estimation and directly performs symbol detection using the (sometimes extremely) limited pilot data. The key innovation in \alg is the proposed decision feedback mechanism in ICL, where we sequentially incorporate the detected symbols into the prompts to improve the detections for subsequent symbols. Extensive experiments across a broad range of wireless communication settings demonstrate that \alg achieves significant performance improvements, in some cases only needing a single pilot pair. 

\end{abstract}


\section{Introduction}
Wireless receiver symbol detection focuses on identifying the transmitted symbols over fading channels with varying signal-to-noise ratios (SNRs). Traditional methods typically follow a two-step process: first estimating the channel using, e.g., the Minimum Mean Square Error (MMSE) estimator, and then performing symbol detection using the estimated channel. However, this approach can be computationally intensive and is impacted by the channel estimation quality. Data-driven approaches, such as deep learning models that directly learn channel estimators and symbol detectors, offer an alternative. Recurrent Neural Networks (RNNs)\cite{lu2018mimo} and Convolutional Neural Networks (CNNs)\cite{neumann2018learning} have been investigated for this task. However, deep neural networks (DNNs) require large amount of data and often perform poorly in the low-data regime. Moreover, adapting pre-trained DNNs to new wireless conditions remains a challenge~\cite{simeone2020learning}.

Advances in Transformer models, particularly decoder-only architectures like GPT \cite{radford2019language}, have demonstrated impressive performance across various fields. Recent result \cite{NEURIPS2020_1457c0d6} shows that pre-trained Transformers can adapt to new tasks during inference through prompt pairs, without requiring explicit model updates. Given an input of the form $(y_1, f(y_1), \ldots, y_n, f(y_n), y_{\text{query}})$, a pre-trained Transformer can approximate $f(y_{\text{query}})$ based on the provided context, where $(y_1, \ldots, y_n, y_{\text{query}})$ represents features and $f$ can represent various classes of functions \cite{garg2022can}.

Wireless symbol detection, which involves estimating transmitted symbols from noisy received signals, aligns well with the Transformer capabilities. \cite{teja2023transformers} introduced Transformers for this task using in-context learning, framing it as a regression problem with MSE loss and achieving near-MMSE performance. Later works expanded this framework: \cite{zecchin2024context} extended it to MIMO systems, and \cite{zecchin2024cell} demonstrated robustness in multi-user MIMO environments. Meanwhile, \cite{abbas2024leveraging} used language models to reformulate detection as a linguistic task. These advances highlight Transformers as a powerful tool for addressing wireless communication challenges.

Despite these successes, prior studies face limitations. Most approaches treat detection as a regression task, requiring MSE-based objectives and post-processing to map continuous outputs to discrete symbols. Additionally, many require ample pilot pairs, which may not be possible in practice, and large models increase inference costs, limiting real-world feasibility.

Inspired by decision feedback in wireless communication (e.g., decision feedback equalizer over multi-path fading channels), we enhance the prompt design by incorporating decision pairs, combining current signals with model detections to improve subsequent detection. Our \alg model uses a carefully designed mixture training process to achieve high performance with limited pilots (sometimes only a single pilot) and maintain accuracy with sufficient pilots. Extensive experiments across modulation schemes validate our approach’s effectiveness.
To summarize, our main contributions include:
\begin{itemize}
    \item We develop a Transformer model that jointly performs channel estimation and symbol detection, departing from prior work that separates these tasks. Our key innovation is the incorporation of decision feedback to improve accuracy and adaptability.
    \item We design a mixed training process that achieves high performance with limited pilots and strong accuracy with abundant pilots, enhancing model practicality for deployment.
    \item We validate our approach with experiments across multiple modulation schemes.
\end{itemize}

\section{System Model and Problem Formulation}

\subsection{Wireless Model}
To more clearly illustrate our design, we consider a canonical receiver symbol detection problem over a standard narrowband wireless fading channel. Specifically, we focus on an $N_r \times N_t$ MIMO system, where the channel is represented by an $N_r \times N_t$ complex-valued matrix $H_t$ at time $t$, following a distribution $P_H$. We normalize the channel coefficients such that each entry in $H_t$ has a unit variance. The received signal at time $t$ is expressed as: $y_t = H_t x_t + z_t,$
where the channel noise $z_t \in \mathbb{C}^{N_r}$ is modeled as a complex  additive white Gaussian noise vector with zero mean and covariance matrix $\sigma^2 I$, following a distribution $P_{\sigma^2}$. Each entry of the input vector $x_t \in \mathbb{C}^{N_t}$ is sampled uniformly at random from a given constellation set $\mathcal{X}$ (e.g., QPSK or 16QAM), and this modulation process is independently and identically distributed (i.i.d.) across both time and space.  We normalize the signal to ensure a unit average total transmit power, i.e. $\mathbb{E}[\|x_t\|^2] = 1$. The average signal-to-noise ratio (SNR) at any receive antenna is given by $\text{SNR} = {1}/{\sigma^2}$.

We focus on the \emph{block-fading} channel model \cite{TV:05} in this paper, where the channel $H_t$ remains constant over a coherent time period of $T$ time slots, and is i.i.d. across different coherence periods. In other words, $H_t=h_l, \forall t=(l-1)T+1,\cdots, lT$ for the $l$-th coherence period where $h_l$ is drawn i.i.d. from $P_H$. Correspondingly, the data transmission is organized into frames, where each frame has a length that is at most $T$. The frame structure is designed such that the first $k$ transmitted symbols are known and pre-determined \emph{pilot symbols}, whose original purposes include assisting the receiver to perform channel estimation of $h_l$ so that it can perform coherent symbol detection. In other words, based on the reception of a few pilot pairs $D_{k} = \{(y_1, x_1), \cdots, (y_k, x_k)\}$, the design goal is to determine a demodulator that accurately recovers the transmitted symbol $x_{k+1}, \cdots, x_{T}$ from the received signal $y_{k+1}, \cdots, y_{T}$ with high probability.

\if{0}
We consider the receiver symbol detection problem over a standard wireless block-fading channel. The channel state remains constant within each frame and is independently and identically distributed (i.i.d.) across frames according to an unknown distribution, as illustrated in \cref{fig:blockfading}. Each frame is divided into pilot signals, which assist in channel estimation, followed by data-carrying received signals.

Specifically, we focus on an $N_r \times N_t$ MIMO system, where the channel is represented by an $N_r \times N_t$ complex-valued matrix $H$, following the distribution $P_H$. The channel noise $z_t \in \mathbb{C}^{N_r}$ is modeled as additive complex Gaussian noise with variance $\sigma^2$, following the distribution $P_{\sigma^2}$. For a given modulation scheme, each entry of the input vector $x \in \mathbb{C}^{N_t}$, representing the transmitted symbol, is uniformly sampled from the modulation constellation set $\mathcal{X}$. We normalize the transmitted signal to ensure an average energy of 1, such that $\mathbb{E}[\|x\|^2] = 1$. The average SNR at any receive antenna is given by: $\text{SNR} = \frac{1}{\sigma^2}$. The received signal is expressed as:
\begin{equation}
    y = Hx + z.
\end{equation}

Within each frame, based on the reception of a few pilot pairs $D_{k} = \{(y_1, x_1), \cdots, (y_k, x_k)\}$, the goal is to determine a demodulator that accurately recovers the transmitted symbol $x_{k+1}$ from the received signal $y_{k+1}$ with high probability.
\fi


\subsection{Existing Methods for Symbol Detection}
In the traditional approach, the receiver first estimates the channel using pilot signals, then performs symbol detection on the received signal $y_{t}$ via hypothesis testing for each $t = k+1, \cdots, T$. Typically, the (Linear) MMSE estimator is used for channel estimation, and the MMSE channel estimate $\hat{H}$ is given by:
$\hat{H} = (X^H X + \sigma^2 I)^{-1} X^H Y,$
where $X$ is the pilot matrix and $Y$ is the received signal matrix. With the estimated channel, the transmitted symbol $\hat{x}_{t}$ is detected by projecting $y_{t}$ onto the closest symbol in the modulation constellation $\mathcal{X}$:
$\hat{x}_{t} = \arg \min_{x \in \mathcal{X}} \| \hat{H}x - y_{t} \|^2, \forall t = k+1, \cdots, T.$

This two-step process treats channel estimation and symbol detection as separate tasks. Such decoupling can result in suboptimal detection, particularly under noisy conditions or limited pilot data. Optimal estimators like MMSE rely on precise statistical models of the channel and noise, which are often hard to obtain. Additionally, these estimators are computationally intensive due to matrix inversions and posterior probability calculations, making them less appealing for real-time applications in high-dimensional systems.

Data-driven, machine learning-based methods\cite{simeone2018very} present a promising alternative. Various neural network architectures, such as deep neural networks (DNNs), convolutional neural networks (CNNs), and recurrent neural networks (RNNs), have been explored to improve channel estimation \cite{le2021deep, neumann2018learning, lu2018mimo, aoudia2021end} and symbol detection \cite{park2020learning, chen2018neural}. While these methods can improve detection, they often suffer from high sample complexity, requiring substantial data for effective training \cite{kim2020massive}. Furthermore, conventional learning models struggle to adapt to varying channel distributions without retraining, limiting their applicability in dynamic wireless environments \cite{simeone2020learning}.

\section{In-Context Learning-Based Symbol Detection}
In-context learning (ICL) for symbol detection leverages the structure of wireless communication frames, particularly in block-fading channels where channel conditions remain stable within each frame. Within these frames, pilot signals are followed by subsequent received signals, which naturally align with the Transformer architecture's strength in processing sequence-based inputs. The Transformer's ability to model dependencies among sequential data allows it to capture complex relationships within transmitted signals, making it highly effective for symbol detection tasks.

In this section, we introduce the ICL-based symbol detection task. First, we provide the formulation of the ICL-based symbol detection problem and then present its implementation using the Transformer model GPT-2, which also serves as the backbone of our proposed solution\footnote{We use GPT-2 with elaborated design choices as a concrete example throughout the paper. However, the proposed principle can be easily adapted to other Transformer architectures.}.

\subsection{Problem Formulation}
\label{sec:problem}
Each ICL detection task $\tau$ corresponds to a latent channel $H$ and a channel noise level $\sigma^2$, following the unknown joint distribution $P_{\tau} = P_H P_{\sigma^2}$. The ICL-based symbol detection does not have prior knowledge of the specific task $\tau$ and is provided only with a prompt $S_{t}^{\tau} = (D_{k}^{\tau}, y_{t})$,
consisting of $k$ target pairs,
$D_{k}^{\tau} = \{(y_1, x_1), \cdots, (y_k, x_k)\}$,
which are sampled from the conditional distribution $P_{x,y|\tau}$ and serve as in-context examples for the current task $\tau$, along with $y_{t}$, $\forall t = k+1, \cdots, T$.

As previously mentioned, each context pair $(x_i, y_i)$ and the inference pair $(y_{t}, x_{t})$ are i.i.d. samples given task $\tau$. For block-fading channels, the context set $D_{k}^{\tau}$, also referred to as pilot data in wireless communication, enhances the channel estimation, thereby improving the reliability and accuracy of data transmission. We note that a significant advantage of the proposed solution is that there is no need to change the frame structure or the design of pilot signals. Rather, the innovation is entirely at the receiver side where we leverage the pilot and decoded signals in a different way. This is an important advantage in practice as it allows for (backward) compatibility with the existing standard.

The goal of symbol detection is to identify the corresponding input signal $x_{t}$ for the new query signal $y_{t}$ from the same task. The ICL-based detection makes its decision as follows:
$\hat{x}_{t} = f_{\theta}(S_{t}^{\tau})$,
where $\theta$ represents the parameters of the model. The detection for the query is measured by the symbol error rate (SER), which is the frequency at which transmitted symbols are incorrectly decoded. The expected SER for the new query with $k$ contexts, taking the expectation over the task distribution for $\forall k = 1, \cdots, T-1$, is defined as:
\begin{equation}
\text{SER}_{k}(\theta) = \mathbb{E}_{\tau}  \mathbb{E}_{x,y|\tau} \left[ f_{\theta}(D_{k}^{\tau}, y_t) \neq x_{t} \right].
\label{eqn:expected_SER}
\end{equation}

\subsection{Vanilla In-Context Symbol Detection}

\begin{figure}[!htbp]
    \vspace{-10pt}  
    \centering
    \includegraphics[width=0.3\textwidth]{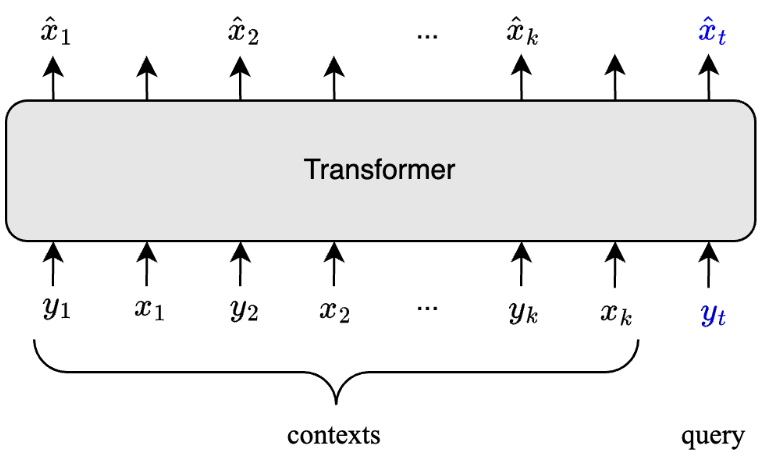}
    \vspace{-10pt} 
    \caption{\footnotesize Decoder-only Transformer architecture for in-context learning-based symbol detection with $k$ pilots. Detection output applies to $\forall t=k+1, \cdots, T$. }
    \vspace{-10pt}  
    \label{fig:model_ICL}
\end{figure}

Transformer models have emerged as powerful tools for symbol detection \cite{shen2024training}, leveraging their ability to capture long-range dependencies for improved detection in varying channels. This approach to wireless symbol detection was introduced in \cite{teja2023transformers, zecchin2024context}. 
The input-output structure is illustrated in \cref{fig:model_ICL}. With a masked self-attention mechanism, the model outputs the detection $\hat{x}_{t}$ at the corresponding position of $y_{t}$, relying only on known preceding contexts and the received signal. During the forward process, the Transformer solves $k+1$ detection problems for the same task $\tau$, using an increasing number of pilot data points. Their results demonstrate that the Transformer exhibits strong capabilities in symbol estimation within context, without requiring explicit model updates.

\section{Decision Feedback IN-Context Detection}
The vanilla ICL approaches for symbol detection require sufficient context to achieve accurate estimation, which is often impractical in real-world scenarios. Pilot signals, essential for these methods, are costly and limited, reducing their adaptability for practical applications. For situations where the amount of pilots is small, neither conventional two-stage (channel estimation then symbol detection) nor vanilla ICL solutions can achieve good performance. Furthermore, these approaches generally formulate the symbol detection task as a regression problem, as shown by \cite{raventos2024pretraining, zecchin2024context, zecchin2024cell}, where a Transformer model was trained to minimize the mean squared error (MSE) loss. Although their model achieved performance comparable to the optimal MMSE estimator for $x$, an additional projection step was required to map the output to the appropriate transmitted symbol, leading to a mismatch and losing optimality in the process.

In contrast, we directly define the problem as a classification task, enabling the model to jointly learn channel estimation and symbol detection while directly measuring the SER during inference. Additionally, we generalize the approach to effectively handle scenarios where pilot information is highly limited by sequentially feeding back the already decoded symbol pairs as \textit{noisy pilots} and incorporating them as part of the prompt. Our model demonstrates robust performance even in challenging conditions with only a \emph{single} pilot, outperforming previous ICL models that struggle with insufficient pilot data. At the same time, it maintains high accuracy when sufficient pilot data is available.

\begin{figure}[!htbp]
    \vspace{-10pt} 
    \centering
    \includegraphics[width=0.3\textwidth]{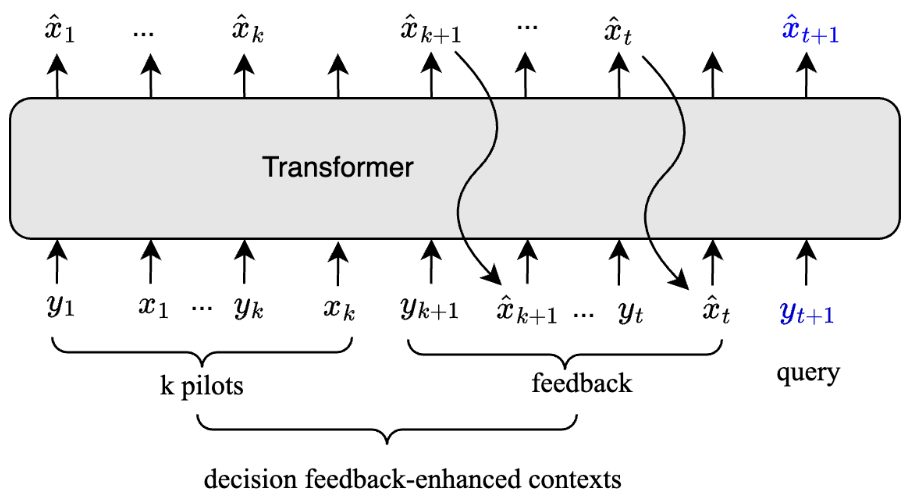}
    \vspace{-10pt} 
 \caption{ \footnotesize \alg model architecture with $k$ pilots and $(t-k)$ decision feedback contexts to detect $x_{t+1}$, $\forall t=k, \cdots, T-1$.}
    \vspace{-5pt} 
    \label{fig:model_DFE}
\end{figure}

Inspired by decision-feedback concepts in wireless communication, we propose the \underline{DE}cision \underline{F}eedback \underline{IN}-Cont\underline{E}xt \underline{D}etection (\alg) model for symbol detection, as shown in \cref{fig:model_DFE}. The \alg model extends the prompt by incorporating the previously received signals and detection decisions alongside prior prompts to improve subsequent detections. While traditional decision-feedback equalizers (DFE) focus on inter-symbol interference (ISI), our study addresses narrowband channels without ISI. Nevertheless, decision feedback is effective here due to the latent common channel. Noisy feedback also provides valuable information, further refining model detection.

\subsection{Model Parameters}

Our specific Transformer model is designed with an embedding dimension of $d_e = 64$, $L = 8$ layers, and $h = 8$ attention heads, resulting in approximately \textbf{0.42 million} parameters, which is significantly smaller compared to large language models (LLMs) commonly applied in wireless communication tasks, such as those discussed by \cite{shao2024wirelessllm, abbas2024leveraging}. For instance, even smaller LLMs like GPT-J 6B contain over 6 billion parameters, making them approximately 14,000 times larger than our model. This compact size not only enables deployment on edge devices but also significantly shortens the inference time, enabling low-latency detection at the receiver. 

\subsection{Training Details}
In this section, we describe our data generation process and the training of the \alg model. Training includes a pre-training phase to equip the model with general predictive abilities and speed up convergence, followed by fine-tuning to adapt the model to scenarios with limited pilot data.

\subsubsection{Data Generation}
We generate data according to the wireless communication model described in \cref{sec:problem}. Specifically, we consider both SISO and 2x2 MIMO systems and explore various modulation schemes, including BPSK, QPSK, 16QAM, and 64QAM. For each wireless system and modulation task, we generate prompts consisting of sequences with $T$ pilot pairs, with the maximum sequence length set to $T = 31$. Both systems operate under a Rayleigh fading channel, where the channel coefficient is sampled as $H \sim \mathcal{CN}(0,1)$. The channel noise is i.i.d. and sampled from a Gaussian distribution, with the noise variance uniformly drawn from the range $[\sigma_{\min}^2, \sigma_{\max}^2]$. The received signal is thus $y_t = H x_t + z_t.$ The training batch size is set to 512.

\subsubsection{ICL Pre-training} 
\begin{figure}[!htbp]
    \centering
    \vspace{-5pt} 
    \includegraphics[width=0.49\textwidth]{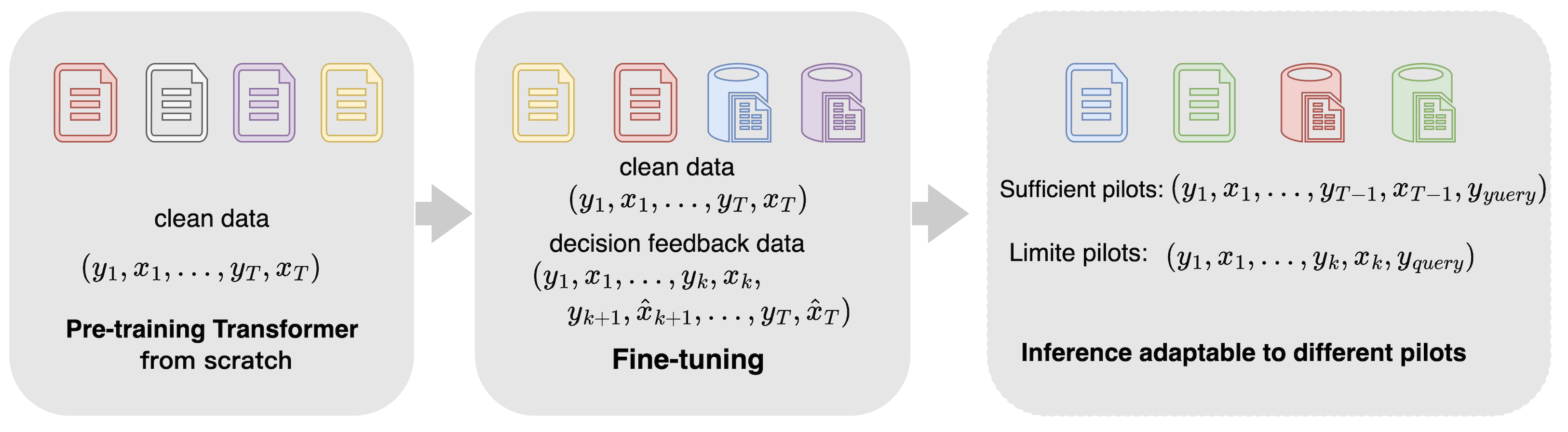}
    \vspace{-20pt} 
    \caption{\footnotesize The training process includes pre-training on clean data, followed by fine-tuning on a mixed dataset of clean and decision feedback noisy data. The model demonstrates strong performance, adapting to both limited and sufficient pilot scenarios during symbol detection (i.e., inference).}
    \label{fig:train_process}
    \vspace{-5pt} 
\end{figure}

We delve into the details of model training, which is divided into two phases, as shown in \cref{fig:train_process}. First, we define ICL training and ICL-testing as operations on ground-truth data, represented by the clean prompt:
$S_{t}^{ICL} = \{y_1, x_1, \ldots, y_{t-1}, x_{t-1}, y_t\}, \quad \text{for } t=1,2,\cdots, T.$
On the other hand, DF-training and DF-testing use iteratively decoded sequences with $k$ pilot data and model decision feedback, which operate on the decision feedback prompt:
$S^{DF}_{t} = \{y_1, x_1, \cdots, y_k, x_k, y_{k+1}, \hat{x}_{k+1}, \ldots, y_{t-1}, \hat{x}_{t-1}, y_t\}, \quad \text{for } t = k+1, \ldots, T,$
where each estimation $\hat{x}_t$ relies on the first $k$ pilot points and prior model decisions.

We define the loss functions for the ICL-training and DF-training models. The adopted loss function is the cross-entropy loss between the model’s output and the ground-truth labels:

\begin{equation}
\mathcal{L}^{ICL}(\theta) = \frac{1}{N T} \sum_{i=1}^{N} \sum_{t=1}^{T}  \text{loss} \left( f_{\theta}(S_{t,i}^{ICL}), x_{t,i} \right),
\label{loss:ICL}
\end{equation}
\vspace{-5pt} 
\begin{equation}    
\mathcal{L}^{DF}(\theta) =  \frac{1}{N T} \sum_{i=1}^{N} \sum_{t=k+1}^{T} \text{loss} \left( f_{\theta}(S_{t,i}^{DF}), x_{t,i} \right),
\label{loss:DF}
\end{equation}
where $\theta$ represents the model parameters, and $N$ is the number of samples.

DF-training  is trained with limited pilot data and utilizes self-sampled labels, where noisy feedback is combined with previous contexts to generate new contexts. This process is time-intensive, as each detection and feedback step requires a model forward pass. Additionally, noisy data complicates convergence, which can cause the model to struggle to converge effectively. Training the Transformer with ICL-training and testing under DF-testing introduces a data mismatch: training uses clean data, while testing involves noise. Despite this mismatch, we observed that although the Transformer’s performance is lower than that achieved with DF-training, it remains highly competitive, suggesting an inherent robustness to noisy data. ICL training is also about ten times faster than DF-training, as it eliminates data sampling and operates solely on clean data.

Considering all factors, our suggested solution is to perform ICL-training first, followed by DF-training. Here, ICL-training serves as pre-training, while DF-training acts as fine-tuning, similar to the pre-training and fine-tuning process used in LLMs. Training epochs are carefully structured into two phases, as shown in \cref{fig:train_process}. In the first phase, the model converges just before reaching a plateau, at which point we transition to the DF-training method. During this transition, a spike in the training loss is observed due to the shift in the training data distribution.  As shown later, ICL pre-training not only accelerates convergence but also improves recognition of clean data and ICL-testing performance.


\subsubsection{Decision Feedback Fine-tuning}

We now elaborate on the details of the carefully designed fine-tuning process, which follows the ICL pre-training phase. The loss function is constructed as a linear combination of the previously defined losses in \cref{loss:ICL,loss:DF}, where $\alpha$ represents the weight of the combination:

\vspace{-5pt}
\begin{equation}
\mathcal{L}^{\text{fine-tuning}}(\theta) = \alpha \mathcal{L}^{DF}(\theta) + (1 - \alpha) \mathcal{L}^{ICL}(\theta).
\vspace{-5pt}
\end{equation}

As explained in the pre-training phase, after ICL pre-training, the model is capable of general symbol detection, performing well on detections with clean data and, to some extent, on detections using the decision feedback method. Furthermore, in the fine-tuning process, the training loss is designed to emphasize decision feedback detection while retaining the model's ability to handle clean data.

Training on both clean and noisy data enhances the robustness of the Transformer model by exposing it to a more diverse dataset. Ultimately, we propose that \textbf{a single Transformer model} can be trained to perform both ICL-testing and DF-testing, making our \alg model adaptable for practical wireless systems. For example, in scenarios with sufficient pilot information, the model can operate in the ICL manner. However, in challenging situations -- common in real-world applications -- where pilots are limited and difficult to acquire, the model can utilize previous decisions to improve performance in subsequent symbol detection.

\section{Experiment}
This section presents the experimental results, analyzing our model's performance against baseline algorithms. Our model excels not only with sufficient pilot data but also shows notable improvement in limited-data scenarios by effectively leveraging noisy feedback. Furthermore, the \alg model demonstrates strong performance in complex modulation tasks, underscoring the Transformer’s capacity to learn geometrical structures within modulation constellations.

\subsection{Baseline Algorithms}
We introduce several baseline algorithms, including the prior ICL model, the MMSE algorithm, and MMSE-DF, a decision-feedback variant of MMSE.

\subsubsection{In-Context Learning}
We train a Transformer using vanilla ICL-training and will plot SER against context sequence length in ICL-testing and DF-testing, shown as the ICL-ICL and ICL-DF lines, respectively. 

\subsubsection{MMSE Algorithm}
We present the MMSE algorithm, assuming a known pilot signal matrix $X$. The received signal matrix is represented as $Y = H X + Z$. Since both the channel and noise follow complex Gaussian distributions, the pair $(H, Y)$ is jointly Gaussian. The MMSE estimator for $H$ is given by:
$\hat{H}_{k}^{\text{MMSE}} = (X^H X + \sigma^2 I)^{-1} X^H Y.$
With the $t$-th received signal $y_{t}$, the transmitted symbol $x_{t}$ is estimated by projection onto the closest symbol in $\mathcal{X}$:
\begin{equation}
\hat{x}_{t} = \arg\min_{x \in \mathcal{X}} \| \hat{H}_{k}^{\text{MMSE}} x - y_{t} \|^2, \forall t = k+1, \cdots, T.
\label{eqn:proejction}
\end{equation}
With $k$ pilot pairs, the mean symbol error rate is computed, shown as a horizontal line labeled MMSE-P$k$.

\subsubsection{MMSE-DF Algorithm}

We introduce MMSE-DF, which uses decision feedback data as a baseline for our \alg model in limited  pilot scenarios. Starting with $k$ pilots, we compute the MMSE estimator of $H$ and detect $\hat{x}_{k+1}$ using $y_{k+1}$ as in \cref{eqn:proejction}. The decision pair $(y_{k+1}, \hat{x}_{k+1})$ merges with the previous dataset, which is then used iteratively to detect each signal until $\hat{x}_{T}$. We plot the SER against the decision feedback-extended context sequence length.

\subsection{Experimental Results}
\begin{figure*}[!htbp]
    \centering
    \begin{minipage}[b]{0.3\textwidth}
        \centering
        \includegraphics[width=\textwidth]{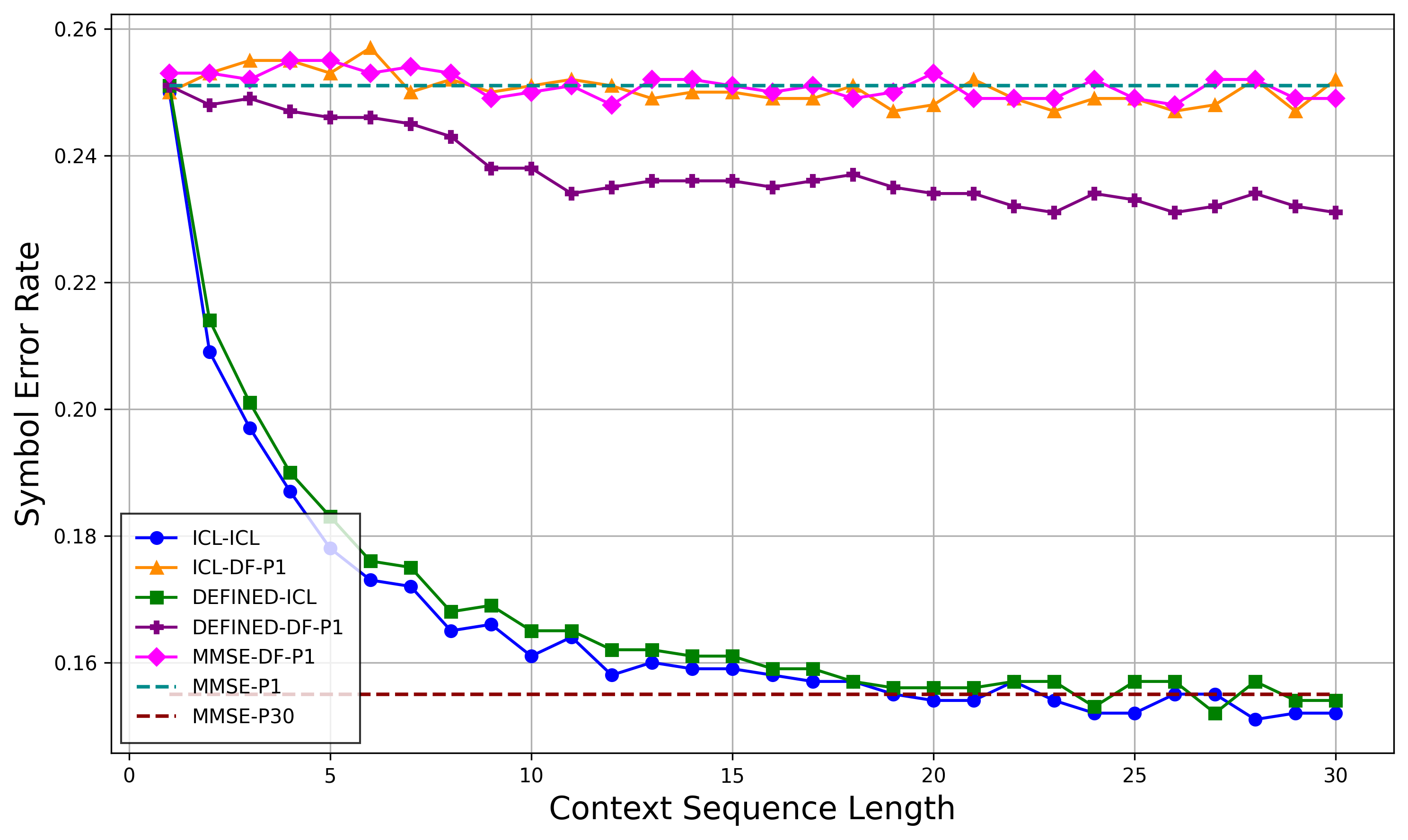}\\
        \small (a) SISO BPSK SNR0 P1\\
        $\text{gain}_{\text{DF}} = 0.076$, $\text{gain}_{\text{ICL}} = 0.380$
    \end{minipage}%
    \hfill
    \begin{minipage}[b]{0.3\textwidth}
        \centering
        \includegraphics[width=\textwidth]{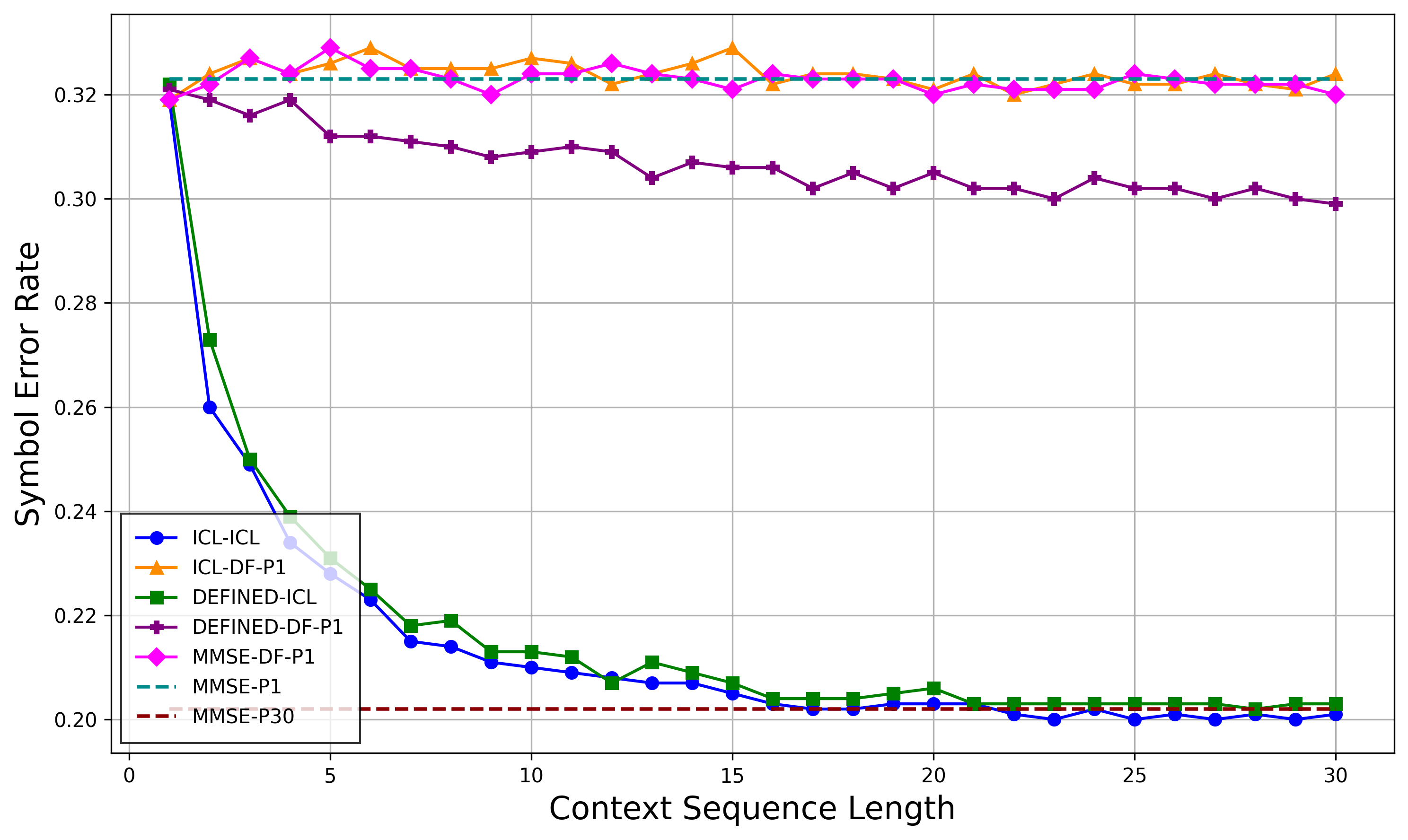}\\
        \small (b) SISO QPSK SNR5 P1\\
        $\text{gain}_{\text{DF}} = 0.056$, $\text{gain}_{\text{ICL}} = 0.372$
    \end{minipage}%
    \hfill
    \begin{minipage}[b]{0.3\textwidth}
        \centering
        \includegraphics[width=\textwidth]{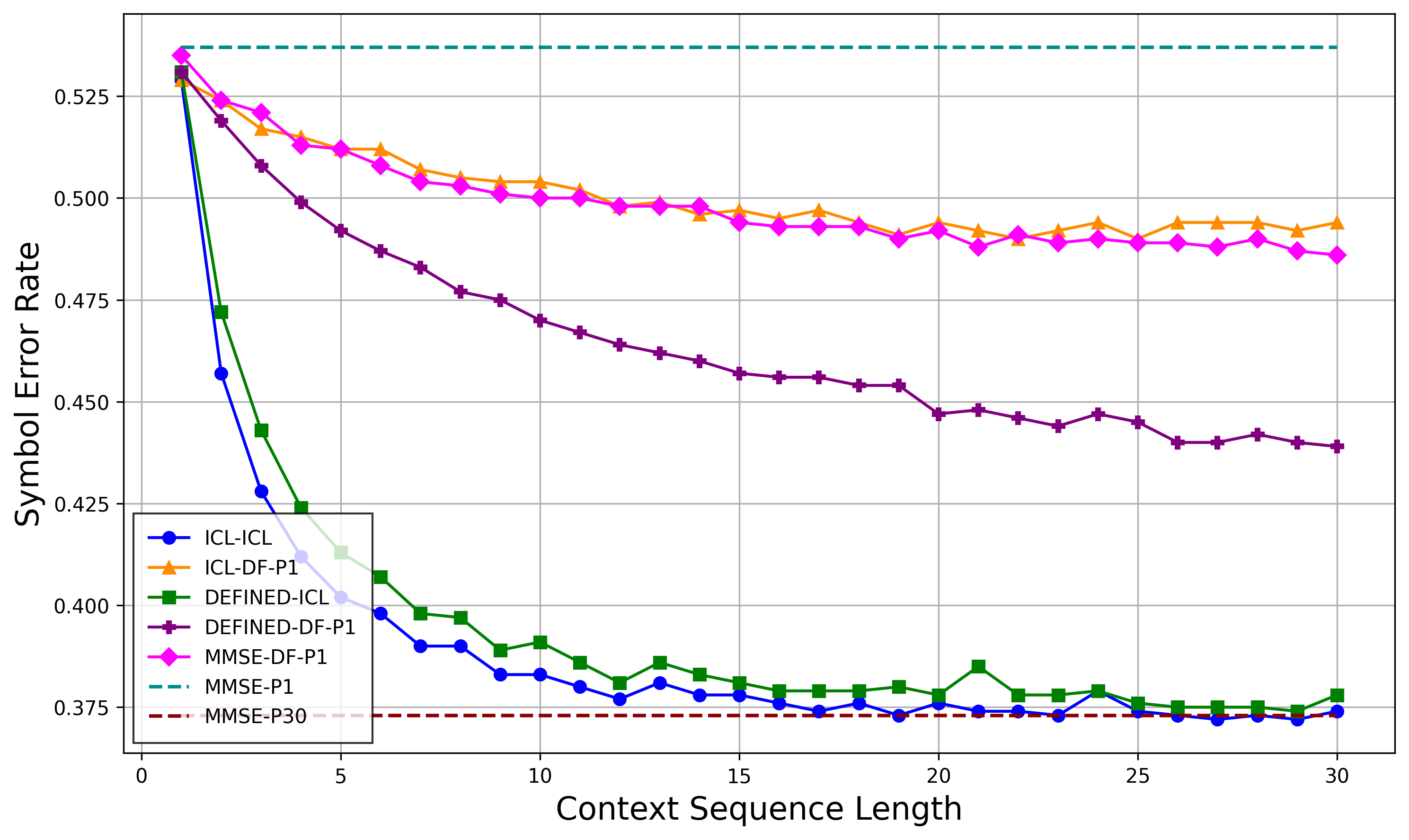}\\
        \small (c) SISO 16QAM SNR10 P1\\
        $\text{gain}_{\text{DF}} = 0.170$, $\text{gain}_{\text{ICL}} = 0.292$
    \end{minipage}
    
    \vspace{1em}
    \begin{minipage}[b]{0.3\textwidth}
        \centering
        \includegraphics[width=\textwidth]{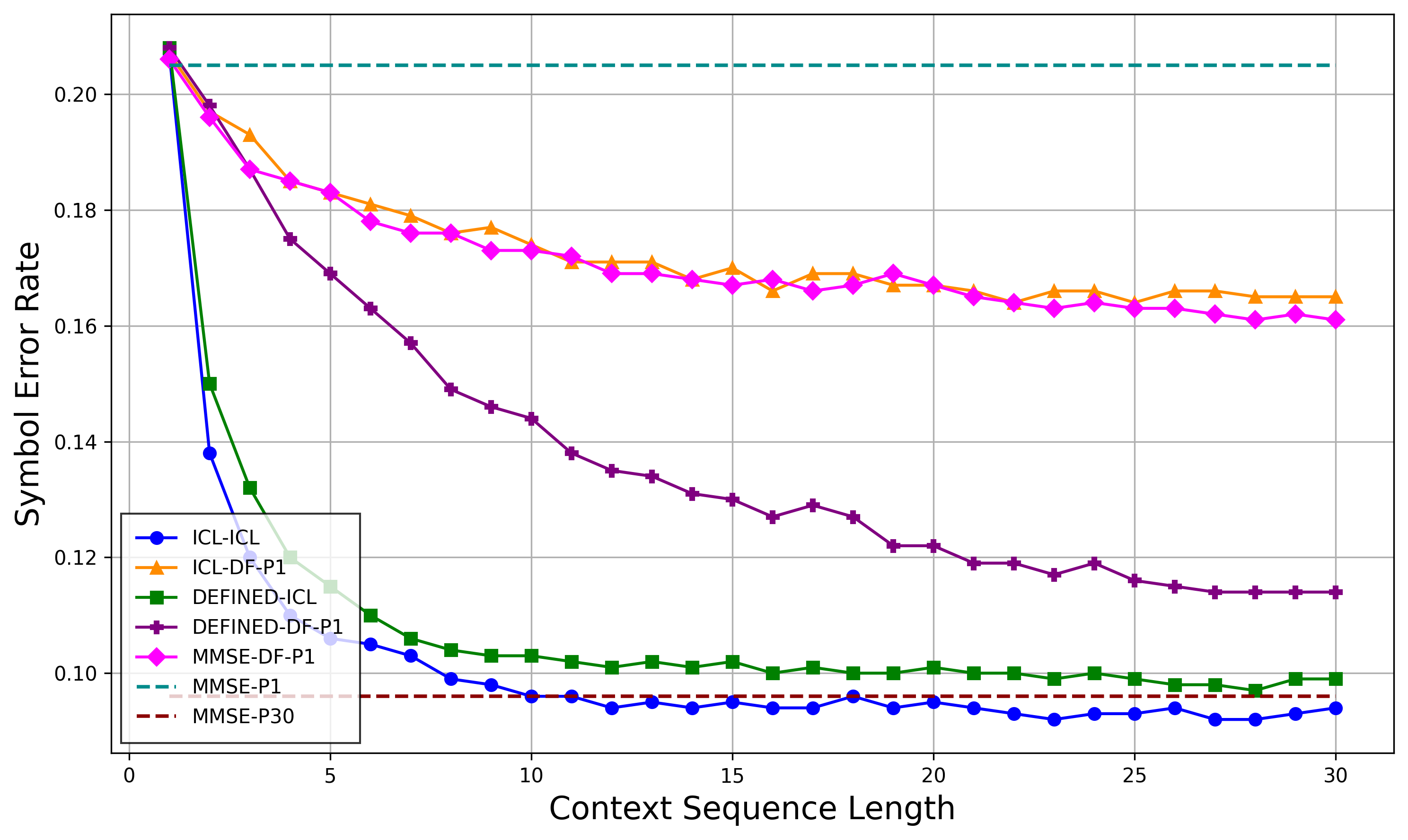}\\
        \small (d) SISO 64QAM SNR25 P1\\
        $\text{gain}_{\text{DF}} = 0.462$, $\text{gain}_{\text{ICL}} = 0.545$
    \end{minipage}%
    \hfill
    \begin{minipage}[b]{0.3\textwidth}
        \centering
        \includegraphics[width=\textwidth]{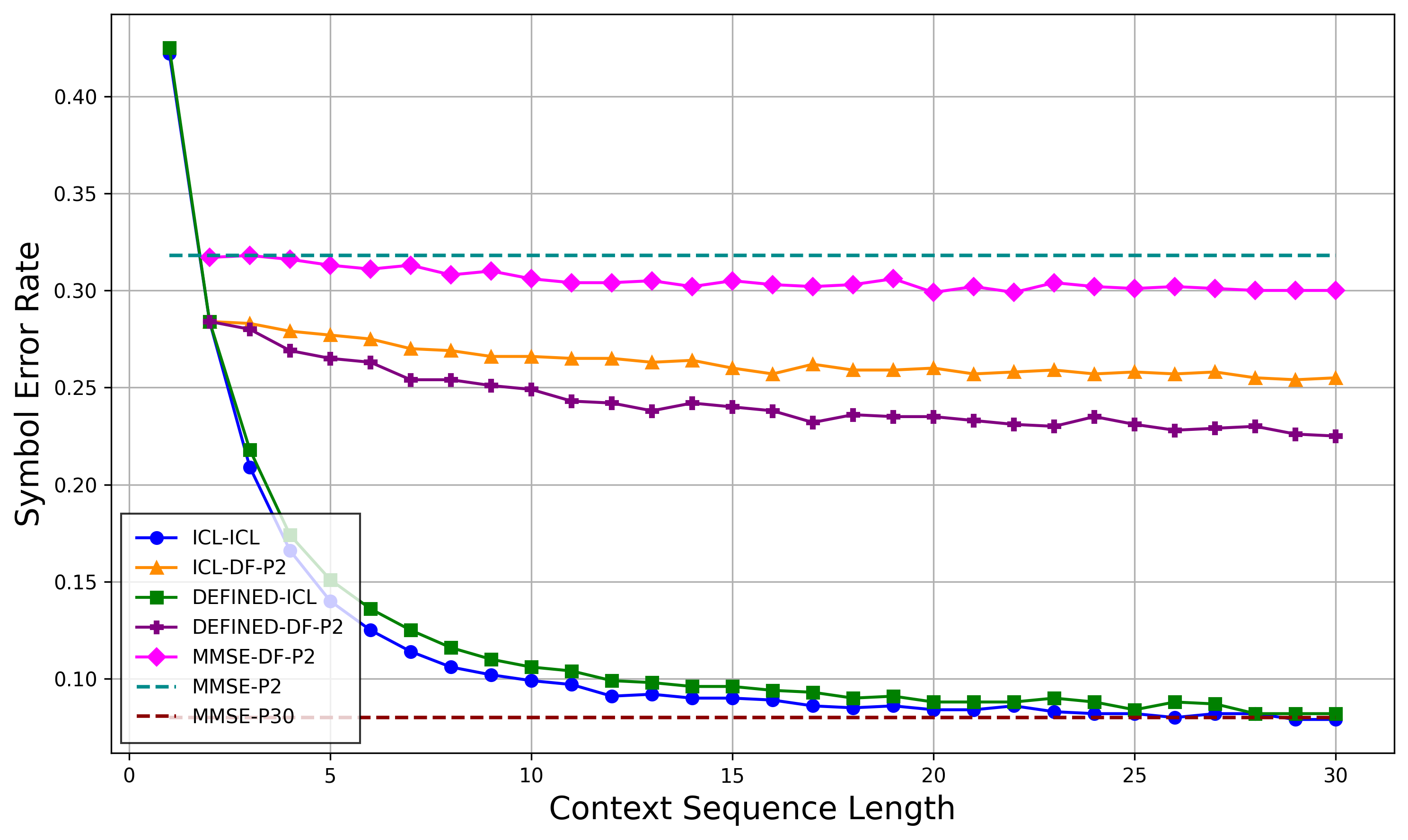}\\
        \small (e) MIMO BPSK SNR5 P2\\
        $\text{gain}_{\text{DF}} = 0.208$, $\text{gain}_{\text{ICL}} = 0.721$
    \end{minipage}%
    \hfill
    \begin{minipage}[b]{0.3\textwidth}
        \centering
        \includegraphics[width=\textwidth]{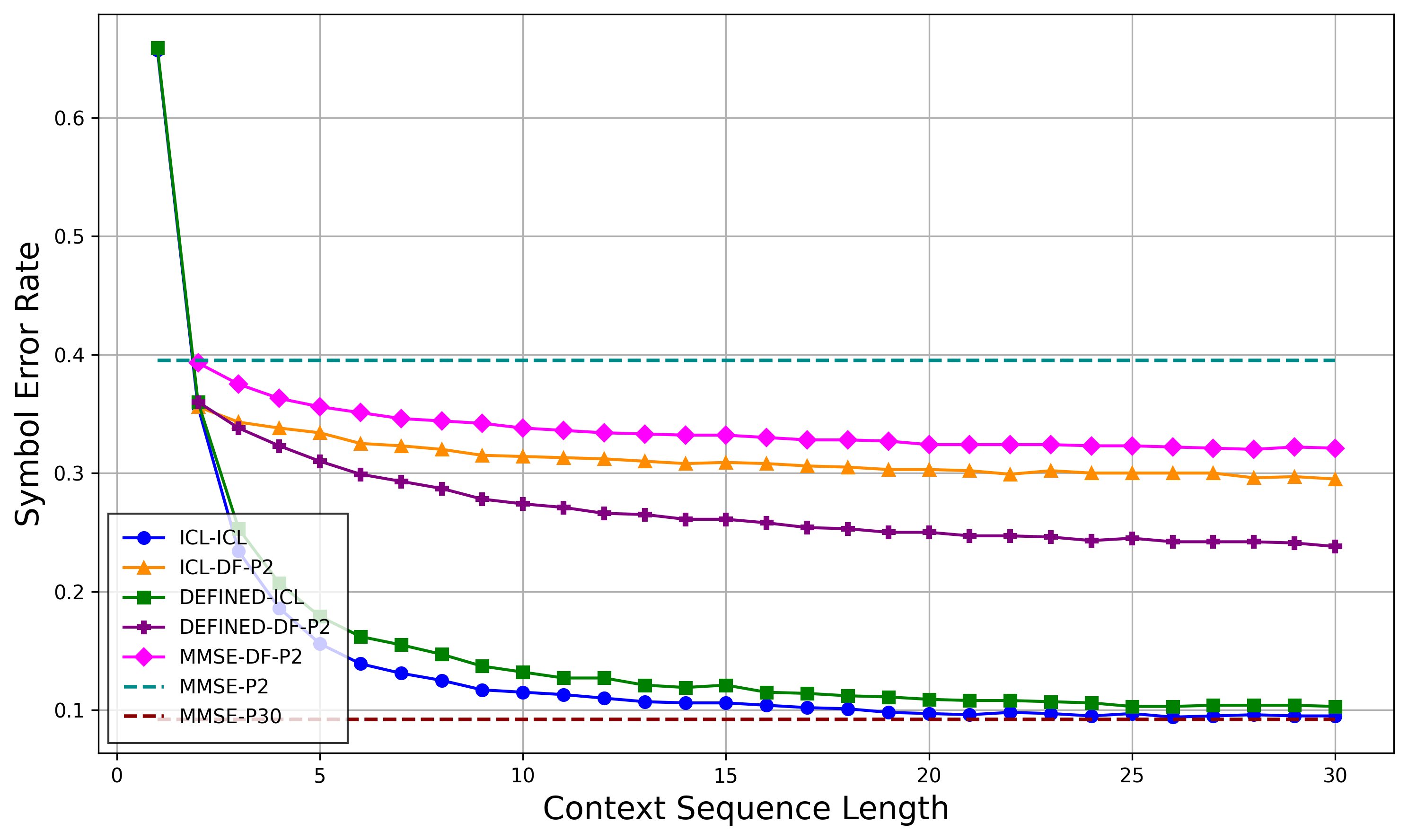}\\
        \small (f) MIMO QPSK SNR10 P2\\
        $\text{gain}_{\text{DF}} = 0.338$, $\text{gain}_{\text{ICL}} = 0.733$
    \end{minipage}
    \caption{\footnotesize SISO performance for BPSK, QPSK, 16QAM, and 64QAM with one pilot, and 2×2 MIMO for BPSK and QPSK with two pilots. The X-axis shows context sequence length, where the $t$-th point for ${*}$-ICL uses $t$ ground truth pilots and ${*}$-DF-P$k$ uses $k$ clean pilots and $(t-k)$ decision feedback noisy pairs.}
    \label{fig:experimental_results}
    \vspace{-10pt} 
\end{figure*}

We set the hyperparameter $\alpha$ to $0.7$ during model training. Testing data is generated with a different random seed, sampling 80,000 prompts to compute the mean SER across tasks. Results are presented for BPSK, QPSK, 16QAM, and 64QAM in the SISO system, and for BPSK and QPSK in a 2x2 MIMO system, plotting SER against context sequence length, as shown in \cref{fig:experimental_results}. The $t$-th point represents the SER of the estimator $\hat{x}_{t+1}$ using $t$ contexts, omitting the 0-th point (random guessing) due to high SER.

To quantify the SER improvement with increasing context length, we define the metric $\text{gain}_{\theta} = \frac{\text{SER}_{k}(\theta) - \text{SER}_{T-1}(\theta)}{\text{SER}_{k}(\theta)}$, representing the relative SER reduction as the context length increases from $k$ to $(T-1)$, starting from $k$ clean pilots. Specifically, $\text{gain}_{DF}$ is computed for our \alg model with the decision feedback-enhanced context sequence, indicating the performance improvement from the decision feedback mechanism, while $\text{gain}_{ICL}$ is computed for the ICL model using the clean pilot sequence.

\subsubsection{Comparison with Baseline Algorithms}

In the plots, the MMSE algorithm, which provides optimal channel estimation based on given pilots and detections of the next transmitted symbol, is shown alongside horizontal lines representing MMSE performance with $k$ pilots and with full (30) pilots, respectively. The ICL-ICL line, for the model trained and tested with ICL method, shows that with 30 pilots, the Transformer slightly outperforms the MMSE algorithm with 30 pilots during ICL testing. This improvement arises from model’s ability to perform channel estimation and symbol detection jointly, leveraging the synergy between these tasks.

The DEFINED-DF line denotes the performance of our proposed model during DF-testing, showing a marked SER reduction as more decision feedback data is incorporated. This confirms that the Transformer can effectively use noisy feedback to improve detections with limited pilot data. Our model also performs well in ICL testing, as indicated by the DEFINED-ICL line, which aligns closely with the ICL-ICL line. This result suggests that ICL pre-training, followed by carefully designed loss functions during decision feedback fine-tuning, allows the model to learn effectively from clean data. Additionally, our \alg model adapts well to real-world symbol detection, excelling with ample pilot data and performing effectively even with a single pilot.

The ICL-DF line, which represents a Transformer trained with ICL but tested under DF conditions, performs significantly worse than our model, nearly coinciding with the MMSE-DF line. This observation highlights that models trained solely on clean data struggle with noisy feedback, producing MMSE-like detections and underscoring the importance of fine-tuning for handling noisy feedback.

\subsubsection{Comparison with Different SNRs and Varying Pilot Lengths}
At high SNR levels, reduced data noise enables more accurate detection from pilot data, enhancing \alg model performance and accentuating the downward SER trend. However, at very high SNRs, the already low initial SER limits further improvement. Our \alg model also performs robustly with minimal pilot data, including the extreme single-pilot cases. As pilot data increases, all algorithms show improved performance in DF inference.

\subsubsection{Comparison of SISO and MIMO Systems}
Across various modulation schemes and SNR levels, our model performs better in MIMO than in SISO systems, with a more pronounced SER reduction from additional decision feedback. This improvement suggests the model effectively learns from the communication system structure and may perform well in multi-user systems.


\subsubsection{Comparison with Different Modulation Schemes}

We conduct experiments using BPSK, QPSK, 16QAM, and 64QAM modulations in the SISO system, representing classification tasks with 2, 4, 16, and 64 classes, respectively. As modulation complexity increases, the detection task becomes more challenging, but we observe greater performance improvements with complex schemes during DF-inference, as reflected by a more pronounced SER decrease with additional feedback data in our \alg model.

Analysis of the Transformer’s output logit vector shows that nearly all of the incorrect detections occur within a small region around the ground-truth label in the constellation. This ``typical error event'' \cite{TV:05} suggests that even incorrect detections carry valuable information, \textit{as the noisy label is often near the correct one, enhancing the model’s detections}. Thus, as modulation complexity increases, the compact constellation set allows noisy feedback to provide more useful information, leading to better SER gains with added feedback data.

These findings demonstrate that our Transformer model effectively captures the constellation set's geometry. It not only learns detections but also recognizes relationships between classes, often assigning higher probabilities to neighboring labels when errors occur. Due to inherent data noise -- e.g., channel fading and additive noise -- received signals with nearby latent labels in the constellation may overlap.
As a result, the model sees adjacent labels as close neighbors, letting its detections retain valuable information, even when they are not entirely accurate.

\section{Conclusion}

Inspired by the decision feedback mechanism in wireless receiver designs, we proposed \alg to enhance symbol detection by incorporating decision pairs into the prompts of Transformer. Our approach achieved significant performance gains with limited pilot data while maintaining high accuracy with sufficient pilot data, demonstrating its adaptability for practical scenarios. Extensive experiments across various modulation schemes validated the robustness and flexibility of our model. These contributions highlight the potential of Transformers, underscoring their capabilities for future wireless communication systems.

\bibliographystyle{IEEEtran}
\bibliography{ref, Shen, wireless}

\end{document}